\documentclass[a4paper,twoside,twocolumn,english,aps,prb]{revtex4}
\usepackage{ae,aecompl}
\usepackage[T1]{fontenc}
\usepackage[latin9]{inputenc}
\usepackage{amsmath}
\usepackage{graphicx}
\usepackage{amssymb}
\usepackage{esint}

\makeatletter

\providecommand{\tabularnewline}{\\}

\@ifundefined{textcolor}{}
{%
 \definecolor{BLACK}{gray}{0}
 \definecolor{WHITE}{gray}{1}
 \definecolor{RED}{rgb}{1,0,0}
 \definecolor{GREEN}{rgb}{0,1,0}
 \definecolor{BLUE}{rgb}{0,0,1}
 \definecolor{CYAN}{cmyk}{1,0,0,0}
 \definecolor{MAGENTA}{cmyk}{0,1,0,0}
 \definecolor{YELLOW}{cmyk}{0,0,1,0}
 }

\usepackage{hyperref}\usepackage{bm}

\makeatother

\usepackage{babel}

\makeatother

\usepackage{babel}

\makeatother

\usepackage{babel}

\makeatother

\usepackage{babel}

\begin{document}

\title{Reptation quantum Monte Carlo for lattice Hamiltonians \\
 with a directed-update scheme}

\author{Giuseppe Carleo, Federico Becca, Saverio Moroni, and Stefano Baroni}

\affiliation{SISSA -- Scuola Internazionale Superiore di Studi Avanzati and \\
 DEMOCRITOS National Simulation Center, Istituto Officina dei Materiali
del CNR \\
 Via Bonomea 265, I-34136, Trieste, Italy}

\date{\today}
\begin{abstract}
We provide an extension to lattice systems of the reptation quantum
Monte Carlo (RQMC) algorithm, originally devised for continuous Hamiltonians.
For systems affected by the sign problem, a method to systematically
improve upon the so-called fixed-node approximation is also proposed.
The generality of the method, which also takes advantage of a canonical
worm algorithm scheme to measure off-diagonal observables, makes it
applicable to a vast variety of quantum systems and eases the study
of their ground-state and excited-states properties. As a case study,
we investigate the quantum dynamics of the one-dimensional Heisenberg
model and we provide accurate estimates of the ground-state energy
of the two-dimensional fermionic Hubbard model. 
\end{abstract}
\maketitle

\section{Introduction}

\label{sec:intro}

The path-integral formulation of quantum mechanics is the foundation
of many numerical methods that allow one to study with great accuracy
the rich physics of interacting quantum systems. At finite temperature,
a path-integral Monte Carlo (PIMC) technique for continuous systems
has been developed and applied by Ceperley and Pollock.\cite{CepPoll,Ceperley:1995}
Recently, this approach has been renovated in a new class of methods
known as {\em worm algorithms}.~\citep{Prokofev:1998,Boninsegni:2006}
The zero-temperature counterparts of the PIMC algorithm are the reptation
quantum Monte Carlo (RQMC)~\citep{Baroni:1999} and the path-integral
ground state (PIGS) methods,~\citep{Sarsa:2000} which have been
demonstrated useful to simulate coupled electron-ion systems,~\citep{Pierleoni}
as well as to infer spectral properties from imaginary time dynamics.~\citep{Carleo:2009}
A number of important physical problems --particularly in the fields
of strongly correlated fermions and cold atoms-- can be fruitfully
modeled by lattice Hamiltonians. A first application of path-integral
techniques to (boson) lattice models was proposed by Krauth {\em
et al.} in 1991.~\citep{Krauth:1991} Few other attempts to apply
PIMC to lattice models have been made ever since, with a recent application
of the RQMC idea to the quantum dimer model Hamiltonian.~\citep{Syljuasen:2006}
In this paper, we propose a new method that generalizes and improves
the approach of Ref.~\onlinecite{Syljuasen:2006} in several ways.
Our method is based on continuous-time random walks and is therefore
unaffected by time-step errors. Inspired by the work of Syljuasen
and Sandvik~\cite{Sandvik:2002} and Rousseau,~\cite{Rousseau:2008}
we adopt a generalization of the {\em bounce algorithm} of Pierleoni
and Ceperley,~\citep{Pierleoni} called {\em directed updates},
which helps reducing the correlation time in path sampling. We also
introduce a worm-algorithm based method to calculate {\em pure}
expectation values of arbitrary off-diagonal observables, which are
generally out of the scope of existing lattice ground-state methods.

The resulting algorithm naturally applies to fermions, using the fixed-node
approximation. A technique to improve systematically upon this approximation
is proposed, based on the calculation of a few moments of the Hamiltonian.
Our methodology is demonstrated by a few case studies on the one-dimensional
Heisenberg and the two-dimensional fermion Hubbard models.

This paper is organized as follow: in Sec.~\ref{sec:formalism} we
present the general formalism of ground-state PIMC for lattice models;
in Sec.~\ref{sec:reptation} our implementation of the RQMC algorithm
on a lattice is presented. In particular, we give a detailed account
of the above mentioned {\em directed update} technique (Sec.~\ref{sub:Directed-updates})
and of the continuous-time propagator (Sec.~\ref{subsec:Contin});
in Sec.~\ref{subsec:OffDiag}, we introduce an extension of the algorithm
to cope with off-diagonal observables, while in Sec.~\ref{subsec:Signpr}
a further extension to systems affected by sign problems is presented,
including a strategy to improve systematically upon the fixed-node
approximation. Sec.~\ref{sec:results} contains a few case applications,
including the simulation of the spectral properties and spin correlations
of the one-dimensional Heisenberg model and the calculation of the
ground-state energies of the fermionic Hubbard model with a significantly
better accuracy than that achieved by the fixed-node approximation.
In Sec.~\ref{sec:conc} we finally draw our conclusions.

\section{General formalism}

\label{sec:formalism}

Let us consider a generic lattice Hamiltonian $\hat{H}$ and a complete
and orthogonal basis set, whose states are denoted by $|x\rangle$.
Given the generic wave function $|\Psi\rangle$, its amplitude on
the configuration $|x\rangle$ will be denoted by $\Psi(x)$, namely
$\Psi(x)=\langle x|\Psi\rangle$. The exact ground-state wave function
$|\Psi_{0}\rangle$ can be obtained by the imaginary time evolution
of a given variational state $|\Psi_{V}\rangle$: \begin{equation}
|\Psi_{0}\rangle\propto\lim_{\beta\rightarrow\infty}|\Psi_{\beta}\rangle,\label{eq:psilim}\end{equation}
 where $|\Psi_{\beta}\rangle\equiv e^{-\beta\hat{H}}|\Psi_{V}\rangle$,
provided that the variational state is non-orthogonal to $|\Psi_{0}\rangle$,
i.e., $\langle\Psi_{V}|\Psi_{0}\rangle\neq0$. Then, the ground-state
expectation value of a quantum operator $\hat{O}$ can be obtained
by \begin{equation}
\langle\hat{O}\rangle=\lim_{\beta\rightarrow\infty}\frac{\langle\Psi_{\beta}|\hat{O}|\Psi_{\beta}\rangle}{\langle\Psi_{\beta}|\Psi_{\beta}\rangle}.\end{equation}

A practical computational scheme can be conveniently introduced by
considering a path-integral representation of the imaginary time evolution.
To such a purpose, we split the total imaginary time $\beta$ into
$M$ slices of {}``duration'' $\tau=\beta/M$, in such a way that
the value of the evolved wave function on a generic many-body state
of the system reads \begin{equation}
\Psi_{\beta}(x_{0})=\sum_{x_{1}\dots x_{M}}\prod_{i=1}^{M}G_{x_{i-1}x_{i}}^{\tau}\Psi_{V}(x_{M}),\label{eq:psibeta}\end{equation}
 where we have introduced the imaginary time propagators \begin{equation}
G_{x_{i-1}x_{i}}^{\tau}=\langle x_{i-1}|e^{-\tau\hat{H}}|x_{i}\rangle.\label{eq:gtau}\end{equation}
 Within this approach, it is easy to write expectation values of operators
$\hat{O}$ that are diagonal in the chosen basis $|x\rangle$, i.e.,
$\langle x|\hat{O}|y\rangle=O(x)\delta_{x,y}$. In fact, in this case
we have that: \begin{equation}
\langle\hat{O}\rangle=\lim_{\beta\rightarrow\infty}\frac{\sum_{\mathbf{X}}\Pi^{\beta}(\mathbf{X})O(x_{M})}{\sum_{\mathbf{X}}\Pi^{\beta}(\mathbf{X})},\label{eq:Obeta}\end{equation}
 where the summation is extended to all possible imaginary time paths
$\mathbf{X}\equiv\{x_{0},x_{1},\dots,x_{2M}\}$, and $\Pi^{\beta}(\mathbf{X})$
is given by: \begin{equation}
\Pi^{\beta}(\mathbf{X})=\Psi_{V}(x_{0})\left[\prod_{i=1}^{2M}G_{x_{i-1}x_{i}}^{\tau}\right]\Psi_{V}(x_{2M}).\label{eq:Pibeta}\end{equation}
 The ground-state energy can be conveniently obtained by means of
the {\em mixed average}: \begin{equation}
E_{0}=\lim_{\beta\rightarrow\infty}\frac{\sum_{\mathbf{X}}\Pi^{\beta}(\mathbf{X})E_{L}(x_{0})}{\sum_{\mathbf{X}}\Pi^{\beta}(\mathbf{X})},\label{eq:energybeta}\end{equation}
 where $E_{L}(x)=\langle x|\hat{H}|\Psi_{V}\rangle/\langle x|\Psi_{V}\rangle$
is the so-called local energy.

Besides the static (i.e., equal-time) correlation functions, this
formalism allows one to calculate also dynamical correlations in imaginary
time $C_{AB}(\mathcal{T})=\langle\hat{A}(\mathcal{T})\hat{B}(0)\rangle$
that can be computed as \begin{equation}
C_{AB}(\mathcal{T})=\lim_{\beta\rightarrow\infty}\frac{\sum_{\mathbf{X}}\Pi^{\beta}(\mathbf{X})A(x_{n})B(x_{m})}{\sum_{\mathbf{X}}\Pi^{\beta}(\mathbf{X})},\label{eq:cabtau}\end{equation}
 where $x_{n}$ and $x_{m}$ are two coordinates of the path such
that $(m-n)\tau=\mathcal{T}$.

\section{Reptation quantum Monte Carlo}

\label{sec:reptation}

A probabilistic interpretation of the previous expectation values~(\ref{eq:Obeta}),~(\ref{eq:energybeta}),
and~(\ref{eq:cabtau}) can be immediately recovered whenever $\Pi^{\beta}(\mathbf{X})\geq0$
for all configurations $\mathbf{X}$. Indeed, in this case, $\Pi^{\beta}(\mathbf{X})$
can be interpreted as a probability distribution that may be readily
sampled by using Monte Carlo algorithms. This fact allows ground-state
expectation values to be calculated exactly, within statistical errors.

\begin{figure}
\includegraphics[width=1\columnwidth]{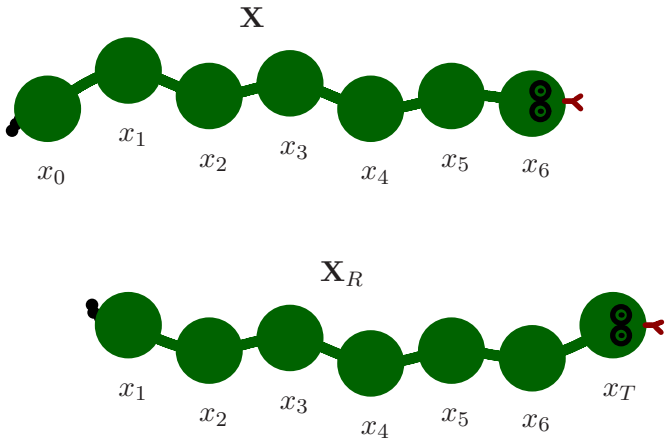} \caption{\label{fig:reptmov} Pictorial representation of the {}``sliding
moves'' along the right imaginary time direction. In the new configuration
(bottom), a new head for the reptile is generated from the old configuration
(top) and the tail is discarded.}

\end{figure}

The basic idea of the RQMC algorithm is to sample the distribution
probability $\Pi^{\beta}(\mathbf{X})$ by using a Markov process with
simple moves. Given the configuration $\mathbf{X}\equiv\{x_{0},x_{1},\dots,x_{2M}\}$,
a new configuration is proposed in two possible ways: either $\mathbf{X}_{L}\equiv\{x_{T},x_{0},\dots,x_{2M-1}\}$
(which we call {}``left move'') or $\mathbf{X}_{R}\equiv\{x_{1},\dots,x_{2M},x_{T}\}$,
(which we call {}``right move''). In both cases, $x_{T}$ is a new
configuration proposed according to a suitable transition probability
$R^{\tau}(x\rightarrow x_{T})$, where $x$ stays for $x_{0}$ ($x_{2M}$)
when the left (right) move is considered. Such {}``sliding moves''
are depicted in Fig.~\ref{fig:reptmov}. Ideally, the transition
probability should guarantee the minimum possible statistical error
on the desired observables and, to such a purpose, it has been proved
useful to consider the propagator with importance sampling, i.e.,
$\tilde{G}_{xy}^{\tau}=G_{xy}^{\tau}\Psi_{V}(y)/\Psi_{V}(x)$ and
take the following transition probability \begin{equation}
R^{\tau}(x\rightarrow y)=\frac{\tilde{G}_{xy}^{\tau}}{w(x)},\label{eq:timpsa}\end{equation}
 where \begin{equation}
w(x)=\sum_{x'}\tilde{G}_{xx'}^{\tau}\end{equation}
 represents the normalization factor. The explicit form of $R^{\tau}(x\rightarrow x_{T})$
will be discussed in more detail in Sec.~\ref{subsec:Contin}. The
proposed configuration $\mathbf{X}_{d}$ (where $d=L$ or $R$) is
accepted or rejected according to the usual Metropolis algorithm,
where the acceptance rate is given by: \begin{equation}
A=\min\left\{ 1\,,\,\frac{\Pi^{\beta}(\mathbf{X}_{d})R^{\tau}(x_{T}\rightarrow x)}{\Pi^{\beta}(\mathbf{X})R^{\tau}(x\rightarrow x_{T})}\right\} .\label{eq:acceptance}\end{equation}
 In this way, a sequence of configurations $\mathbf{X}^{k}$ is generated,
$k$ being the (discrete) time index of the Markov chain.

%


In order to reduce the auto-correlation time of the observables it
is convenient to make several consecutive sliding moves along the
same imaginary time direction.~\cite{Baroni:1999} To such a purpose,
a recent development called {}``bounce'' algorithm has been proposed.~\cite{Pierleoni}
Although the bounce algorithm sampling procedure does not fulfill
the microscopic detailed balance, the equilibrium probability $\Pi^{\beta}(\mathbf{X})$
is correctly sampled.~\cite{Pierleoni} The RQMC algorithm with bounce
moves can be then summarized in the following steps: 
\begin{enumerate}
\item For the current direction of the move and for the present configuration
$\mathbf{X}^{k}$, propose $x_{T}$ according to the transition probability
$R^{\tau}(x\rightarrow x_{T})$, where $x=x_{0}^{k}$ if $d=L$ and
$x=x_{2M}^{k}$ if $d=R$. 
\item Given the form of the acceptance ratio $A$ of Eq.~(\ref{eq:acceptance}),
accept the proposed configuration according to the probability \begin{equation}
A_{L}=\min\left\{ 1\,,\,\frac{w(x_{0}^{k})}{w(x_{2M-1}^{k})}\right\} ,\label{eq:accL}\end{equation}
 if $d=L$, or with probability \begin{equation}
A_{R}=\min\left\{ 1\,,\,\frac{w(x_{2M}^{k})}{w(x_{1}^{k})}\right\} ,\label{eq:accR}\end{equation}
 if $d=R$. 
\item If the move is accepted, update the path configurations according
to $\mathbf{X}^{k+1}=\mathbf{X}_{d}$ and continue along the same
direction, otherwise $\mathbf{X}^{k+1}=\mathbf{X}^{k}$ and change
direction. 
\item Go back to 1. 
\end{enumerate}

\subsection{Directed updates}

\label{sub:Directed-updates}

At this point we introduce a novel alternative sampling approach,
which generalizes the bounce idea while strictly fulfilling the detailed
balance condition. Such a scheme, which is largely inspired by the
{\em loop algorithm} methods devised for the stochastic series
expansion~\citep{Sandvik:2002,Sandvik:1999} and for the {\em worm
algorithm},~\citep{Rousseau:2008,Rombouts:2006} allows one to choose
the time direction in a purely Markovian way, i.e., independently
of the previous history.

In our algorithm, a Markov step consists of many simple consecutive
{}``sliding moves'', whose number is not fixed {\em a-priori}
but is determined by a certain probability (see below). The actual
Monte Carlo step takes place at the end of few consecutive updates
along the currently chosen direction. In the examples below, we denote
the number of these sliding moves between two Monte Carlo steps by
$s$.

At the beginning of each Markov step we choose a direction $d$ according
to the probability $P(\mathbf{X}^{k},d)$, whose form will be specified
later. Assuming that the right direction has been chosen, we propose
a new configuration $x_{T}$, according to the transition probability
$R^{\tau}(x_{2M}^{k}\rightarrow x_{T})$ and the configuration labels
are shifted according to $\mathbf{X}^{k+1}=\{x_{1}^{k},\dots,x_{2M}^{k},x_{T}\}$,
with $x_{2M}^{k+1}=x_{T}$. At this point, we continue updates along
this direction with probability $K(\mathbf{X}^{k+1},\rightarrow)$,
or stop with probability $[1-K(\mathbf{X}^{k+1},\rightarrow)]$. If
it has been decided to continue the updates, then a new configuration
is generated according to $R^{\tau}(x_{2M}^{k+1}\rightarrow x_{T})$
and the labels of the configuration are again shifted according to
$\mathbf{X}^{k+2}=\{x_{1}^{k+1},\dots,x_{2M}^{k+1},x_{T}\}$. The
Markov step finishes after $s$ consecutive updates along the right
direction only when $K(\mathbf{X}^{k+s},\rightarrow)<\xi_{s}$, where
$\xi_{s}$ is a random number uniformly distributed in $[0,1)$. At
this point a Metropolis test should be done, in order to accept or
reject the sequence of intermediate $s$ sliding moves: \begin{equation}
A=\min\left\{ 1,\frac{q(\mathbf{X}^{k+s})}{q(\mathbf{X}^{k})}\right\} ,\label{eq:acceptdir}\end{equation}
 where (see Appendix~\ref{append}) \begin{eqnarray}
q(\mathbf{X}) & = & \frac{P(\mathbf{X},\leftarrow)}{1-K(\mathbf{X},\rightarrow)}w(x_{2M-1})\nonumber \\
 & = & \frac{P(\mathbf{X},\rightarrow)}{1-K(\mathbf{X},\leftarrow)}w(x_{1}).\end{eqnarray}
 However, in order to avoid time-consuming restorations of the original
configuration, it is preferable to accept all the moves, while keeping
track of the residual weight $q(\mathbf{X})$. This is possible since
$A$ only depends upon initial and final configurations, so that,
given that all the intermediate moves are accepted, the sampled distribution
probability is $\Pi^{\beta}(\mathbf{X})\times q(\mathbf{X})$. The
contribution of the current configuration to statistical averages
must be then weighted by the factor $1/q(\mathbf{X})$.

To proceed to the next Markov step, a new direction $d$ is chosen
according to $P(\mathbf{X}^{k+s},\leftarrow)$ and $P(\mathbf{X}^{k+s},\rightarrow)$
and the updates are carried along the extracted new direction.

Let us now show the actual expressions for the afore-mentioned probabilities.
In Appendix~\ref{append}, it is demonstrated that the detailed balance
is satisfied if one chooses the probabilities for the directions as
\begin{eqnarray}
P(\mathbf{X},\leftarrow) & = & \frac{1}{1+a(\mathbf{X})},\label{eq:Pleft}\\
P(\mathbf{X},\rightarrow) & = & \frac{a(\mathbf{X})}{1+a(\mathbf{X})},\label{eq:Prig}\end{eqnarray}
 where \begin{equation}
a(\mathbf{X})=\frac{w(x_{2M-1})}{w(x_{1})}\frac{1-K(\mathbf{X},\leftarrow)}{1-K(\mathbf{X},\rightarrow)},\end{equation}
 which is positive and, therefore, guarantees that the above defined
quantities are well defined probabilities, i.e., $0\leq P(\mathbf{X},\leftarrow)\leq1$
and $0\leq P(\mathbf{X},\rightarrow)\leq1$, with the additional property
that $P(\mathbf{X},\leftarrow)+P(\mathbf{X},\rightarrow)=1$.

Regarding the probabilities to continue the updates along the current
direction, we have a substantial freedom of choice, provided that
the condition $\frac{K(\mathbf{X},\leftarrow)}{K(\mathbf{X},\rightarrow)}=\frac{w(x_{1})}{w(x_{2M-1})}$,
is satisfied (see Appendix~\ref{append}). In this paper we have
adopted \begin{eqnarray}
K(\mathbf{X},\leftarrow) & = & \alpha\min\left\{ 1,b(\mathbf{X})\right\} ,\label{eq:pkeeplefA}\\
K(\mathbf{X},\rightarrow) & = & \alpha\min\left\{ 1,\frac{1}{b(\mathbf{X})}\right\} ,\label{eq:pkeeprigA}\end{eqnarray}
 where we have defined: \begin{equation}
b(\mathbf{X})=\frac{w(x_{1})}{w(x_{2M-1})}.\end{equation}
 and $0<\alpha<1$ is an arbitrary parameter of the algorithm, which
controls the average number of consecutive updates along the same
direction.

Summarizing, the RQMC algorithm with directed updates consists of
a sequence of Markov steps determined by the following rules: 
\begin{enumerate}
\item Choose a time direction $d$ according to the probabilities of Eqs.~(\ref{eq:Pleft})
and~(\ref{eq:Prig}). 
\item Propose a new configuration $x_{T}$ according to the transition probability
$R^{\tau}(x\rightarrow x_{T})$, where $x=x_{0}^{k}$ if $d=L$ and
$x=x_{2M}^{k}$ if $d=R$. 
\item Shift the configuration indexes according to $\mathbf{X}^{k+1}=\{x_{T},x_{0}^{k},\dots,x_{2M-1}^{k}\}$
if $d=L$ or $\mathbf{X}^{k+1}=\{x_{1}^{k},\dots,x_{2M}^{k},x_{T}\}$
if $d=R$. 
\item According to the probability $K(\mathbf{X}^{k},\rightarrow)$ or $K(\mathbf{X}^{k},\leftarrow)$,
decide whether keep moving in the same direction or change direction.
In the former case, go to 2, otherwise go to 5. 
\item The Markov step ends here and the current configuration carries the
weight $1/q(\mathbf{X}^{k+s})$, where $s$ is the number of intermediate
moves along the direction chosen. 
\end{enumerate}
The relationship between the directed update scheme and the bounce
algorithm is further elucidated in Appendix~\ref{sec:On-the-relationship},
where general considerations about the efficiency of the algorithms
are presented.

\subsection{Continuous-time propagator}

\label{subsec:Contin}

One of the most striking differences between the original formulation
of the RQMC on the continuum and the present formulation on the lattice
is the lack of the discretization error appearing in the Trotter decomposition
of the propagator. Indeed it is easier to carry the propagation in
continuous imaginary time on a lattice,~\citep{Capriotti:2000} than
on the continuum.~\citep{Lee:2005} To such a purpose, let us consider
the limit of an infinitesimal imaginary time $\epsilon$, for which
the transition probability of Eq.~(\ref{eq:timpsa}) can be written
as \begin{eqnarray}
 &  & R^{\epsilon}(x\rightarrow y)\simeq\frac{\delta_{xy}-\epsilon\Psi_{V}(y)H_{xy}/\Psi_{V}(x)}{1-\epsilon E_{L}(x)}\\
 &  & \simeq\delta_{xy}\left[1+\epsilon E_{L}(x)\right]-\epsilon\left[H_{xy}\frac{\Psi_{V}(y)}{\Psi_{V}(x)}\right]+o(\epsilon^{2}),\label{eq:timpsast}\end{eqnarray}
 where $E_{L}(x)$ is the previously defined local energy and $H_{x,y}=\langle x|H|y\rangle$
denotes the matrix element of the Hamiltonian. Whenever $\Psi_{V}(y)H_{xy}/\Psi_{V}(x)$
is {\em non positive} for all $x$ and $y$, this equation takes
the form of a continuous-time Markov process, whose analytical properties
are well known. In particular, the probability distribution for the
{}``waiting time'' $\tau_{w}$ in a given state $x$, i.e., the
average time that the system spends in the state $x$ before making
an off-diagonal transition to another state $y\neq x$, is exactly
known, namely $P(\tau_{w};x)=\exp\{-\tau_{w}\left[H_{xx}-E_{L}(x)\right]\}$.
As a consequence, the finite-time propagator $R^{\tau}(x\rightarrow y)$
can be directly sampled, giving rise to a succession of a certain
number $n$ of consecutive transitions $x\rightarrow z_{1}\rightarrow z_{2}\rightarrow\dots\rightarrow y$,
with corresponding waiting times $\tau_{w}(z_{i})$ (such that $\sum_{i}\tau_{w}(z_{i})=\tau$).
The normalization of the whole process is \begin{equation}
w(x)=\exp\left[-\sum_{i}\tau_{w}(z_{i})E_{L}(z_{i})\right],\label{eq:ctw}\end{equation}
 where the waiting times are extracted according to the exponential
probability $P(\tau_{w};z_{i})$. The transitions between the intermediate
configurations are done according to the off-diagonal elements of
Eq.~(\ref{eq:timpsast}), i.e., $z_{i+1}$ is chosen with probability
proportional to $\left[-\Psi_{V}(z_{i+1})H_{z_{i}z_{i+1}}/\Psi_{V}(z_{i})\right]$.

\subsection{Off-diagonal observables}

\label{subsec:OffDiag}

The formalism so-far developed allows one to successfully compute
{\em pure} ground-state expectation values of operators that are
{\em diagonal} in the local basis $x$, with the expectation values
of off-diagonal operators restricted to the so-called {\em mixed
averages}.~\cite{Baroni:1999,Sarsa:2000,Capriotti:2000} Nonetheless,
it is often of great interest to remove such a limitation (whose result
is biased by the quality of the variational wave function) and a dedicated
sampling strategy has to be devised in order to cope with such a need.
In the following, we show that a relatively easy modification of the
sampling scheme can accomplish this task, providing us with a general
tool to compute ground-state averages of operators that are non local
in the chosen basis $x$.

Let us consider an arbitrary off-diagonal observable $\hat{\mathcal{O}}$,
which can be in turn considered as the summation of many observables
we are interested in, i.e., $\hat{\mathcal{O}}=\sum_{d}\hat{\mathcal{O}}^{(d)}$.
For example, we can imagine these operators to be the components of
the one-body density matrix at a given distance, $\hat{\mathcal{O}}^{(d)}=\sum_{\left\langle r,r^{\prime}\right\rangle _{d}}b_{r}^{\dagger}b_{r^{\prime}}$
with the summation extended to all lattice coordinates at a fixed
distance $d$.

In the spirit of Refs~\onlinecite{Rombouts:2006,Rousseau:2008}
we introduce a {\em worm-operator} defined by \begin{equation}
\mathcal{W}_{x,y}=\delta_{x,y}+\lambda\mathcal{O}_{x,y},\label{eq:wormopera}\end{equation}
 where $\lambda$ is a positive constant, and consider the extended
configuration space spanned by the paths \begin{eqnarray}
\Pi_{\mathcal{W}}^{\beta}(\mathbf{X}) & = & \Psi_{V}\left(x_{0}\right)\times\prod_{i=1}^{L}G_{x_{i-1}x_{i}}^{\tau}\times\mathcal{W}_{x_{L}x_{R}}\times\nonumber \\
 & \times & \prod_{i=R+1}^{2M+1}G_{x_{i-1}x_{i}}^{\tau}\times\Psi_{V}\left(x_{2M+1}\right).\label{eq:PiWorm}\end{eqnarray}
 The extended paths are broken in two (space)-discontinuous pieces
by the worm operator, which is placed at an imaginary-time $0\leq\tau_{LR}\leq\beta$.
Therefore, paths contain $2(M+1)$ coordinates, including $x_{L}$
and $x_{R}$ that refer to the same imaginary time $\tau_{LR}$.

The configuration space spanned by Eq.~(\ref{eq:PiWorm}) is clearly
larger than the one spanned by Eq.~(\ref{eq:Pibeta}), which is recovered
whenever $x_{L}=x_{R}$, i.e., when the worm operator is {\em diagonal}.

\begin{figure}
\includegraphics[width=1\columnwidth]{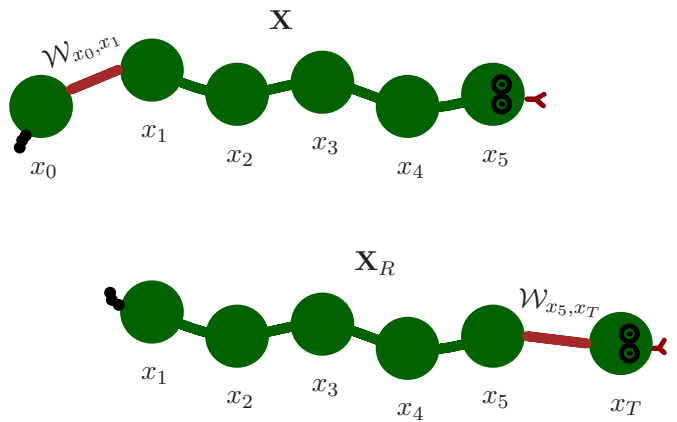} \caption{\label{fig:reptmovworm} Pictorial representation of the {}``sliding
moves'' along the right imaginary-time direction when the worm operator
sits at the tail of the reptile. In the new configuration (bottom),
a new head for the reptile is generated from the old configuration
(top), the old tail configuration is discarded and the worm discontinuity
is moved to the {}``neck'' of the reptile.}

\end{figure}

The pure ground-state expectation value of the operator $\hat{\mathcal{O}}$
is conveniently written in terms of the extended paths as \begin{equation}
\langle\hat{\mathcal{O}}\rangle=\frac{1}{\lambda}\lim_{\beta\rightarrow\infty}\frac{\sum_{\mathbf{X}}\Pi_{\mathcal{W}}^{\beta}(\mathbf{X})\times\Theta(x_{L}\neq x_{R})}{\sum_{\mathbf{X}}\Pi_{\mathcal{W}}^{\beta}(\mathbf{X})\times\Theta(x_{L}=x_{R})},\label{eq:offdiag}\end{equation}
 where $\Theta(C)\neq0$ whenever condition $C$ is satisfied. The
modulus of Eq.~(\ref{eq:PiWorm}) can be in turn interpreted as a
probability distribution and stochastically sampled by means of the
elementary sliding moves considered before. Indeed, whenever the worm
operator is far from the ends of the imaginary-time paths, the sampling
scheme remains unchanged. In this case, a move along direction $d$
will generate a new head (or tail) for the reptile according to $R^{\tau}(x\rightarrow x_{T})$
while shifting the worm position of $\pm\tau$. On the other hand,
whenever the worm operator reaches the ends of the reptile, a new
worm configuration is proposed on the other side; in analogy with
the previous analysis, new configurations are generated according
to a transition probability \begin{equation}
R^{\mathcal{W}}(x\rightarrow y)=\frac{1}{\bar{w}(x)}\left|\mathcal{W}_{xy}\frac{\psi_{V}(y)}{\psi_{V}(x)}\right|,\label{eq:wormpsa}\end{equation}
 where $\bar{w}(x)$ is the normalization factor. Due to the particular
form of the matrix elements~(\ref{eq:wormopera}), the transition
probability will lead either to diagonal configurations $(x=y)$ or
to off-diagonal configurations $(x\neq y)$, thus generating continuous
and discontinuous paths. The relative probability for diagonal and
off-diagonal configurations depends on the value of $\lambda$ that
can be tuned in order to reach a balanced sampling frequency for the
different sectors of the extended paths. In order to exemplify the
worm updates, let us consider the case in which $d=R$ and a configuration
$\Psi_{V}\left(x_{0}\right)\mathcal{W}_{x_{0}x_{1}}\left[\prod_{i=2}^{2M+1}G_{x_{i-1}x_{i}}^{\tau}\right]\Psi_{V}\left(x_{2M+1}\right)$,
after a sliding update in the right direction, we will have $\Psi_{V}\left(x_{1}\right)\left[\prod_{i=2}^{2M+1}G_{x_{i-1}x_{i}}^{\tau}\right]\mathcal{W}_{x_{2M+1}x_{T}}\Psi_{V}\left(x_{T}\right)$,
where $x_{T}$ is proposed according to the transition probability
$R^{\mathcal{W}}(x_{2M+1}\rightarrow x_{T})$ (see Fig.~\ref{fig:reptmovworm}).
In analogy with the previous case, the acceptance factor for the bounce
moves reads $\bar{A}_{R}=\min\left\{ 1\,,\,\frac{\bar{w}(x_{2M+1})}{\bar{w}(x_{1}^{k})}\right\} $.

%


Summarizing, the RQMC with worm-updates consists of the following
steps: 
\begin{enumerate}
\item For the current direction of the move $d$ and for the present configuration
$\mathbf{X}^{k}$ consider the worm-operator position $\tau_{LR}$. 
\item If the worm is not at the ends of the reptile (i.e., $\tau_{LR}\neq0$
when $d=L$ and $\tau_{LR}\neq\beta$ when $d=R$) go to step (a),
otherwise go to step (b).

\begin{enumerate}
\item Propose a new configuration $x_{T}$ according to the transition probability
$R^{\tau}(x\rightarrow x_{T})$, where $x=x_{0}^{k}$ if $d=L$ and
$x=x_{2M+1}^{k}$ if $d=R$. The new configuration is accepted with
probability \begin{equation}
A_{L}=\min\left\{ 1\,,\,\frac{w(x_{0}^{k})}{w(x_{2M}^{k})}\right\} ,\label{eq:accL-1}\end{equation}
 if $d=L$, or with probability \begin{equation}
A_{R}=\min\left\{ 1\,,\,\frac{w(x_{2M+1}^{k})}{w(x_{1}^{k})}\right\} ,\label{eq:accR-1}\end{equation}
 if $d=R$. In the proposed state $\mathbf{X}_{d}$, all the configuration
labels are shifted in the $d$ direction, determining in turn a shift
of the worm operator of a time interval $\pm\tau$, depending on $d$. 
\item Propose a new configuration $x_{T}$ according to the worm transition
probability $R^{\mathcal{W}}(x\rightarrow x_{T})$, where $x=x_{L}^{k}=x_{0}^{k}$
if $d=L$ and $x=x_{R}^{k}=x_{2M+1}^{k}$ if $d=R$. Accept the new
configuration with probability \begin{equation}
\bar{A}_{L}=\min\left\{ 1\,,\,\frac{\bar{w}(x_{0}^{k})}{\bar{w}(x_{2M}^{k})}\right\} ,\label{eq:accLbar}\end{equation}
 if $d=L$, or with probability \begin{equation}
\bar{A}_{R}=\min\left\{ 1\,,\,\frac{\bar{w}(x_{2M+1}^{k})}{\bar{w}(x_{1}^{k})}\right\} ,\label{eq:accRbar}\end{equation}
 if $d=R$. In the proposed state $\mathbf{X}_{d}$, all the configuration
labels are shifted in the $d$ direction, and the worm operator is
moved from the head (tail) to the tail (head) of the reptile, depending
on $d$. 
\end{enumerate}
\item If the move is accepted, update the path configurations according
to $\mathbf{X}^{k+1}=\mathbf{X}_{d}$ and continue along the same
direction, otherwise $\mathbf{X}^{k+1}=\mathbf{X}^{k}$ and change
direction. 
\item Go back to 1. 
\end{enumerate}
This scheme samples the probability density associated to the modulus
of Eq.~(\ref{eq:PiWorm}), and the expectation values of the individual
components $\mathcal{\hat{O}}_{d}$ can be recast as statistical averages
over such a probability distribution, while keeping track of the overall
sign of the extended paths. In particular the best estimate of the
ground-state expectation values is obtained when the worm is in the
central part of the path, at $\tau_{LR}=\beta/2$, leading to \begin{eqnarray}
 &  & \langle\hat{\mathcal{O}}^{(d)}\rangle=\frac{\sum_{\mathbf{X}}\Pi_{\mathcal{W}}^{\beta}(\mathbf{X})\times\Theta\left(\mathcal{O}_{x_{L}x_{R}}^{(d)}\neq0,\,\tau_{LR}=\frac{\beta}{2}\right)}{\sum_{\mathbf{X}}\Pi_{\mathcal{W}}^{\beta}(\mathbf{X})\times\Theta\left(x_{L}=x_{R},\,\tau_{LR}=\frac{\beta}{2}\right)}\nonumber \\
 &  & =\frac{1}{\lambda}\frac{\left\langle \Theta\left(\mathcal{O}_{x_{L}x_{R}}^{(d)}\neq0\right)\times\text{sign}\left[\Pi_{\mathcal{W}}^{\beta}(\mathbf{X})\right]\right\rangle _{\text{OD}}^{\text{center}}}{N_{D}^{\text{center}}},\label{eq:ostatoffdiag}\end{eqnarray}
 where $\left\langle \dots\right\rangle _{\text{OD}}^{\text{center}}$
denotes statistical averages over the off-diagonal distribution $\left|\Pi_{\mathcal{W}}^{\beta}(\mathbf{X})\right|\Theta(x_{L}\neq x_{R},\,\tau_{LR}=\frac{\beta}{2})$
and $N_{D}^{\text{center}}$ is the number of configurations sampled
with a diagonal worm operator in the center of the paths.

\subsection{Tackling the sign problem}

\label{subsec:Signpr}

When the probability distribution of Eq.~(\ref{eq:Pibeta}) is not
positive defined, as is generally the case with fermions, the probabilistic
interpretation of the imaginary time paths breaks down. This circumstance,
which is known as {}``sign problem'', originates whenever $\Psi_{V}(y)H_{xy}/\Psi_{V}(x)>0$
for some element $x\ne y$. In this case, it is not possible to have
polynomial algorithms that are able to obtain an {\em exact} solution
of the problem, which would imply to sample correctly the resulting
signs. Therefore, approximated schemes are welcome and often adopted,
the most widespread one being the so-called fixed-node (FN) approximation.
For lattice systems, this approach relies on the definition of an
effective Hamiltonian, which depends parametrically on the nodal structure
of a {\em given} variational wave function $\Psi_{V}(x)=\langle x|\Psi_{V}\rangle$.~\cite{CeperleyFN}
The matrix elements of the FN Hamiltonian are defined as \begin{equation}
H_{xy}^{\text{{fn}}}=\begin{cases}
H_{xx}+\nu_{\text{sf}}(x) & \text{if}\,\, x=y\\
H_{xy} & \text{if}\,\,\Psi_{V}(y)H_{xy}\Psi_{V}(x)\leq0\\
0 & \text{if}\,\,\Psi_{V}(y)H_{xy}\Psi_{V}(x)>0\end{cases}\label{eq:hfnmel}\end{equation}
 where the sign-flip potential is $\nu_{\text{sf}}(x)=\sum_{y:\text{sf}}\Psi_{V}(y)H_{xy}/\Psi_{V}(x)$,
the sum being extended to all the sign-flip states defined by the
condition $\Psi_{V}(y)H_{xy}\Psi_{V}(x)>0$. With such a choice, the
transition matrix $R^{\epsilon}(x\rightarrow y)$ of Eq.~(\ref{eq:timpsast})
is always positive definite and the summation of Eq.~(\ref{eq:psibeta})
is now restricted --which results in the FN approximation-- to a region
of the Hilbert space in which imaginary time paths are positive definite.
Therefore, within the FN approximation, the ground-state wave function
$|\Psi^{\text{fn}}\rangle$ of $\hat{H}^{\text{fn}}$ can be stochastically
sampled without any sign problem. Moreover, it is easy to show that
the FN approximation becomes exact whenever the signs of the exact
ground state are known. Most importantly, it has been proven~\cite{CeperleyFN}
that the FN ground-state energy $E^{\text{fn}}=\langle\hat{H}^{\text{fn}}\rangle$
gives a rigorous upper-bound to the exact ground-state one and improves
the pure variational results.

At this point, we introduce a straightforward, although computationally
expensive, way to improve further the FN energy. Our strategy amounts
to compute the expectation values of arbitrary powers of the original
Hamiltonian $\hat{H}$ on the FN ground state $|\Psi_{\text{fn}}\rangle$,
namely \begin{equation}
L_{k}=\frac{\langle\Psi_{\text{fn}}|\hat{H}^{k}|\Psi_{\text{fn}}\rangle}{\langle\Psi_{\text{fn}}|\Psi_{\text{fn}}\rangle}.\label{eq:Lk}\end{equation}
 The FN ground state can be expanded in the basis set of the eigenstates
of $\hat{H}$ as $|\Psi_{\text{fn}}\rangle=\gamma_{0}|\Psi_{0}\rangle+\gamma_{1}|\Psi_{1}\rangle+\gamma_{2}|\Psi_{2}\rangle+\dots$
and $L_{k}=\gamma_{0}^{2}E_{0}^{k}+\gamma_{1}^{2}E_{1}^{k}+\gamma_{2}^{2}E_{2}^{k}+\dots$,
with $\sum_{i}\gamma_{i}^{2}=1$.

Since very often the FN wave function has a considerable overlap with
only few low-energy states, the knowledge of the first few moments
of the Hamiltonian are enough to approximately reconstruct both the
coefficients $\gamma_{i}$ and the energies $E_{i}$. To such a purpose,
let us consider a typical situation in which only the first $2n$
moments of the Hamiltonian have been numerically calculated and are
therefore known. We can then truncate the expansion for $L_{k}$ to
the order $n-1$ having a closed system of $2n$ equations \begin{equation}
L_{k}=\sum_{i=0}^{n-1}{\gamma}_{i,n}^{2}{E}_{i,n}^{k},\label{eq:truncL}\end{equation}
 for $k=0,\dots2n-1$ that can be solved for the unknowns $\gamma_{i,n}$
and ${E_{i,n}}$. In the limit of large $n$, the approximated $E_{0,n}$
converges to the exact ground-state energy. Moreover, we verified
that $E_{0,n}\geq E_{0}$, as a result of a connection between the
solutions of the Eq.~(\ref{eq:truncL}) and the Lanczos procedure
written in terms of the moments of the Hamiltonian.~\cite{Whitehead:1978}

The Hamiltonian moments are off-diagonal operators and can, in principle,
measured according to the sampling procedure detailed in Sec.~\ref{subsec:OffDiag}.
In the present implementation we are able to achieve sufficient statistical
accuracy only for the first moment of the Hamiltonian, i.e., $L_{1}=\langle\hat{H}\rangle$,
while higher moments are too noisy. Yet, to our knowledge our algorithm
is the only one that allows the calculation of the expectation value
of the {\em original} Hamiltonian $\hat{H}$. This is known \cite{CeperleyFN}
to be a better upper bound than the expectation value of the FN Hamiltonian
accessible with other zero-temperature algorithms.

Although we are not currently in position to measure directly the
Hamiltonian moments $L_{k}$ we have a controlled access to the {\em
mixed averages} \begin{equation}
L_{k}^{\text{mix}}=\frac{\langle\Psi_{\text{fn}}|\hat{H}^{k}|\Psi_{V}\rangle}{\langle\Psi_{\text{fn}}|\Psi_{V}\rangle},\label{eq:Lkmix}\end{equation}
 which present optimal statistical uncertainty. Moreover, an improved
estimate of the ground-state energy based on the knowledge of the
first few moments $L_{k}^{\text{mix}}$ can be obtained solving a
system of equation similar to Eq.~(\ref{eq:truncL}) that leads to
the approximate ground-state energies ${E_{i,n}^{\text{mix}}}$. Unfortunately,
the proof that ${E_{i,n}^{\text{mix}}}\geq E_{0}$, for $n>1$, is
far from being trivial, requiring a generalization of the already
non-trivial upper bound for $n=1$ described in Ref.~\onlinecite{CeperleyFN}.
Nonetheless, we have numerically verified that, in all the cases treated
in this paper (where $E_{0}$ is {\em a-priori} known), the condition
${E_{i,n}^{\text{mix}}}\geq E_{0}$ is always verified. We are then
led to conjecture that this may always be the case.

\section{Results}

\label{sec:results}

\subsection{Low-energy excitations and spin correlations of the Heisenberg model}

Hereafter, we present a simple application of the previous ideas to
sign-problem free spin Hamiltonians. Let us consider the one-dimensional
quantum Heisenberg model \begin{equation}
\hat{H}=J\sum_{i}\hat{{\bf S}}_{i}\cdot\hat{{\bf S}}_{i+1},\end{equation}
 where $\hat{{\bf S}}_{i}=\left(\hat{S}_{i}^{x},\hat{S}_{i}^{y},\hat{S}_{i}^{z}\right)$
is the spin 1/2 operator on the site $i$ and $J>0$ is the nearest-neighbor
super-exchange coupling.

%
\begin{figure}
\includegraphics[width=1\columnwidth]{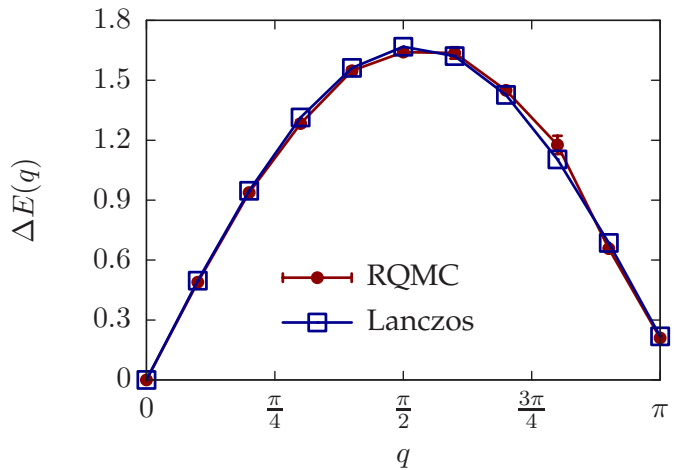} \caption{\label{fig:Sq20} Lowest-energy excitations as a function of the wave-vector
$q$ for an $L=20$ Heisenberg chain. The energies are extracted from
the dynamical structure factor $S(q,\omega)$ and are compared to
exact results by the Lanczos method.}

\end{figure}

\begin{figure}
\includegraphics[width=1\columnwidth]{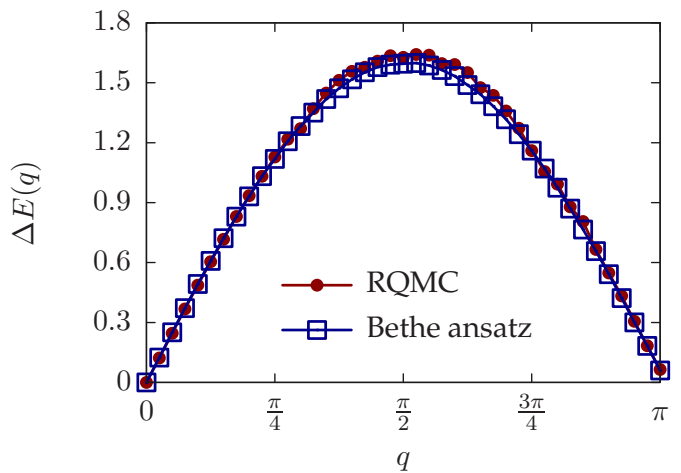} \caption{\label{fig:Sq80} The same as in Fig.~\ref{fig:Sq20} for $L=80$.
Exact results are given by Bethe ansatz.}

\end{figure}


The total number of sites is denoted by $L$ and periodic-boundary
conditions are assumed. This model can be solved exactly by using
the so-called Bethe ansatz technique.~\cite{bethe} Information on
the excitation spectrum can be obtained from the dynamical structure
factor \begin{equation}
S(q,\omega)=\int dt\langle\hat{S}_{q}^{z}(t)\hat{S}_{-q}^{z}(0)\rangle e^{i\omega t},\end{equation}
 where $\hat{S}_{q}^{z}(t)=1/\sqrt{L}\sum_{j}\hat{S}_{j}^{z}(t)e^{iqj}$
is the Fourier transform of time-evolved spin projection on the z-axis.
By introducing a complete set of eigenstates of the Hamiltonian $|\Psi_{n}\rangle$
with eigenvalues $E_{n}$, we have that \begin{equation}
S(q,\omega)=\sum_{n\ne0}|\langle\Psi_{0}|\hat{S}_{q}^{z}|\Psi_{n}\rangle|^{2}\delta(\omega-\omega_{n}),\label{eq:sqomega}\end{equation}
 where $\omega_{n}=(E_{n}-E_{0})$. In the thermodynamic limit, the
spin-1 states form a branch, which is very similar to spin waves in
standard ordered systems, although no long-range order is found in
one dimension.

Imaginary time correlation functions of arbitrary (diagonal) operators
can be efficiently evaluated via Eq.~(\ref{eq:cabtau}). This fact
allows us to have a direct access to $S(q,\mathcal{T})=\langle\hat{S}_{q}^{z}(\mathcal{T})\hat{S}_{-q}^{z}(0)\rangle$.
This imaginary time correlation function can be then analytically
continued, by using the Maximum-Entropy method,~\citep{Gubernatis:1991}
in order to have a reasonable numerical estimate for the dynamical
structure factor of Eq.~(\ref{eq:sqomega}).

Before presenting the results, let us mention that we consider the
following Jastrow state as a variational wave function:~\cite{Manousakis:1991,Franjic:1997}
\begin{equation}
|\Psi_{V}\rangle=\exp\left[\sum_{i,j}v_{ij}\hat{S}_{i}^{z}\hat{S}_{j}^{z}\right]|FM\rangle\end{equation}
 where $|FM\rangle$ is the ferromagnetic state along the x direction,
for which $\langle x|FM\rangle$ does not depend upon the spin configuration
and the variational parameters $v_{ij}$ are optimized by using the
method of Ref.~\onlinecite{Sorella:2005}.

In Fig.~\ref{fig:Sq20}, we show the results for a small $L=20$
system, where exact diagonalizations are possible by using the Lanczos
method. We report the energy excitations $\Delta E(q)=E_{q}-E_{0}$
for the lowest state with $S=1$ and fixed momentum $q$. In this
case a perfect agreement between our RQMC results and the exact ones
is found. Moreover, also on larger systems a very good accuracy is
possible (see Fig.~\ref{fig:Sq80}), demonstrating the performances
of our numerical algorithm.%
\begin{figure}
\includegraphics[width=1\columnwidth]{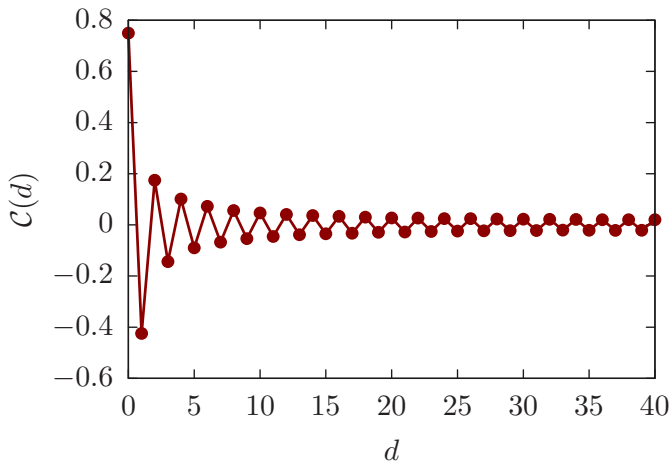} \caption{\label{fig:corrheis} Ground-state expectation value of the spin-spin
correlation function $\mathcal{C}(d)$ for the Heisenberg model on
a $80$-site chain. }

\end{figure}

In order to exemplify the potentialities of the scheme outlined in
\ref{subsec:OffDiag}, we conclude this part of the results devoted
to the Heisenberg model showing the ground-state expectation value
of the spin-spin correlation at distance $d$ \begin{eqnarray}
\mathcal{C}(d) & = & \frac{1}{L}\sum_{i}\left(\hat{{\bf S}}_{i}\cdot\hat{{\bf S}}_{i+d}\right).\label{eq:cofd}\end{eqnarray}
 The desired observable is used as a worm operator and the value of
the correlation function at the various distances is computed by means
of the estimator of Eq. \eqref{eq:ostatoffdiag}. In Fig. \ref{fig:corrheis},
we show the expectation value of $\mathcal{C}(d)$ for a $80$-site
one-dimensional lattice. In this case we are able to achieve very
good statistics for the off-diagonal observable, with a relatively
negligible computational effort, when compared to the evaluation of
the ground-state expectation value of other diagonal observables.

\subsection{Ground-state properties of the fermionic Hubbard model}

As an example of the application of the RQMC to sign-problem affected
Hamiltonians, we present some results for the fermionic Hubbard model
on a square lattice, defined by: \begin{equation}
\hat{H}=-t\sum_{\langle i,j\rangle,\sigma}\hat{c}_{i,\sigma}^{\dagger}\hat{c}_{j,\sigma}+h.c.+U\sum_{i}\hat{n}_{i,\uparrow}\hat{n}_{i,\downarrow},\end{equation}
 where $\langle\dots\rangle$ indicate nearest-neighbor sites, $\hat{c}_{i,\sigma}^{\dagger}$
($\hat{c}_{i,\sigma}$) creates (destroys) an electron on the site
$i$ with spin $\sigma$, and $\hat{n}_{i,\sigma}=\hat{c}_{i,\sigma}^{\dagger}\hat{c}_{i,\sigma}$.
As a variational state we consider \begin{equation}
|\Psi_{V}\rangle=\exp\left[\sum_{i,j}v_{ij}\hat{n}_{i}\hat{n}_{j}\right]|FS\rangle\end{equation}
 where $|FS\rangle$ is the non-interacting Fermi sea and the Jastrow
factor involves density-density correlations. The variational parameters
$v_{ij}$ entering in the Jastrow factor may be optimized again by
minimizing the variational energy with the method of Ref.~\onlinecite{Sorella:2005}.
In order to avoid open shells in $|FS\rangle$, we consider $45$-degrees
tilted lattices with $L=2\times l^{2}$ sites, such that both the
half-filled case and selected holes-doped cases are closed shells.
\begin{figure}[b]
 \includegraphics[width=1\columnwidth]{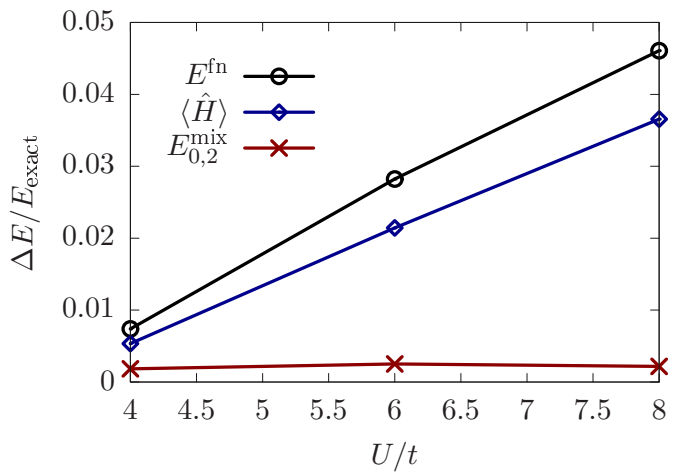}
\caption{\label{fig:18sites} Ground-state energy for the fermionic Hubbard
model at half filling on a $18$-sites tilted-square lattice. The
energy difference $\Delta E=E_{\text{exact}}-E$ is computed with
distinct approximations described in the text.}

\end{figure}


Let us start by showing the results for $18$ electrons on $18$ sites,
where Lanczos diagonalizations are possible.~\cite{Becca:2000} In
Fig.~\ref{fig:18sites}, we report our results for the ground-state
energy. The FN approach gives rather accurate results for small values
of $U/t$, i.e., $U/t\lesssim4$, where $(E_{\text{exact}}-E^{\text{fn}})/E_{\text{exact}}\lesssim0.01$.
By increasing the on-site interaction, the FN approach becomes worse
and worse. This fact is due to the choice of the variational wave
function that does not contain antiferromagnetic order. Remarkably,
a considerable improvement may be obtained by considering the {\em
pure} expectation value of the Hamiltonian, which is systematically
lower than the FN energy, as demonstrated in Ref.~\onlinecite{CeperleyFN}
and now accessible within our algorithm. Further improvements to the
FN energy can be obtained upon considering few (up to three) higher
moments of the Hamiltonian measured as mixed-averages, see Fig.~\ref{fig:18sites}.
The scheme based upon the Hamiltonian moments (described in Sec.~\ref{subsec:Signpr})
allows us to reach a great accuracy for the ground state energy, with
a residual error almost independent of $U/t$. Indeed, in this way
we have $(E_{\text{exact}}-E)/E_{\text{exact}}\lesssim0.002$ up to
$U/t=8$.

This approach remains very effective also for larger systems, even
though the variational wave function loses accuracy by increasing
the cluster size (because the ground state has antiferromagnetic order
in the thermodynamic limit, while the variational state is paramagnetic).
In Table~\ref{tab:50sites}, we report the ground-state energy for
$50$ sites for the half-filled case, while in Table~\ref{tab:50sitesholes}
we report the ground-state energies for selected cases at finite hole-doping,
where numerically exact results (for moderate values of $U$ and moderate
lattice sizes) can be obtained by the Auxiliary-Field Monte Carlo
method.~\cite{SorellaPriv}

\begin{table}
\begin{tabular}{cccc}
\hline 
$U/t$  & $E^{\text{fn}}$  & $\langle\hat{H}\rangle$  & $E_{0,2}^{\text{mix}}$ \tabularnewline
\hline
\hline 
$4$  & $-42.850(1)$  & $-43.16(1)$  & $-43.282(1)$\tabularnewline
$5$  & $-36.364(1)$  & $-36.51(1)$  & $-37.052(1)$\tabularnewline
$6$  & $-31.885(1)$  & $-32.17(1)$  & $-32.640(1)$\tabularnewline
$7$  & $-28.318(1)$  & $-28.66(1)$  & $-29.022(1)$\tabularnewline
$8$  & $-25.382(1)$  & $-25.62(1)$  & $-26.056(1)$\tabularnewline
\end{tabular}\caption{\label{tab:50sites} Ground-state energy as a function of the Hubbard
$U$ repulsion on the $50$-site lattice at half filling.}

\end{table}

\begin{table}
\begin{tabular}{ccccc}
\hline 
$N$  & $E^{\text{fn}}$  & $\langle\hat{H}\rangle$  & $E_{0,2}^{\text{mix}}$  & $E_{AF}$ \tabularnewline
\hline
\hline 
$50$  & $-42.850(1)$  & $-43.16(1)$  & $-43.282(1)$  & $-43.983(1)$\tabularnewline
$42$  & $-53.402(1)$  & $-53.57(1)$  & $-53.769(1)$  & $-54.001(1)$\tabularnewline
$26$  & $-55.4325(1)$  & $-55.63(1)$  & $-55.6112(1)$  & $-55.782(1)$\tabularnewline
$18$  & $-50.4127(1)$  & $-50.50(1)$  & $-50.4383(1)$  & $-50.474(1)$\tabularnewline
\end{tabular}\caption{\label{tab:50sitesholes} Ground-state energy as a function of the
number of electrons $N$ for Hubbard repulsion $U/t=4$ on a $50$-site
lattice. The numerically exact results obtained by the Auxiliary-Field
Monte Carlo method $E_{AF}$ are also shown for comparison.~\cite{SorellaPriv}}

\end{table}

\section{Conclusions}

\label{sec:conc}

In this paper we have provided an efficient and general formulation
of the reptation quantum Monte Carlo technique on lattice models.
In particular, we showed an alternative sampling approach which generalizes
the bounce algorithm, previously introduced to reduce auto-correlation
time of the observables. Our scheme allows one to choose the time
direction in a purely Markovian way. In addition, the average number
of consecutive moves along the time directions may be optimized by
a fine tuning of a certain parameter that has been expressly introduced
in the transition probabilities. We reported benchmarks for two different
models with pure bosonic and fermionic degrees of freedom, by showing
to what extent it is possible to have accurate results both on the
ground state and low-energy excitations. The introduction of a general
method to compute ground-state expectation values of arbitrary off-diagonal
observables also constitutes an important achievement, which will
ease the study of relevant properties such as Bose-Einstein condensation
and superconductivity phenomena in strongly interacting models. In
addition, the possibility to directly measure the {\em pure} ground-state
expectation values may open the way to a better optimization of the
correlated wave function associated to the ground-state of an effective
Hamiltonian which is not the FN one.

\acknowledgments It is a pleasure to acknowledge here precious discussions
with S. Sorella and A. Parola. We also acknowledge support from CINECA
and COFIN 07.

\appendix

\section{Derivation of the probabilities for the directed-update scheme}

\label{append}

In this Appendix we give a detailed derivation of the probabilities
for the directed updates. The detailed balance condition guarantees
that the given probability distribution $\Pi^{\beta}(\mathbf{X})$
is sampled if transitions from an initial state $\mathbf{X}^{k}$
to a final state $\mathbf{X}^{k+s}$ differing for $s$ intermediate
updates are accepted according to: \begin{equation}
A^{s}=\min\left\{ 1,\frac{\Pi^{\beta}(\mathbf{X}^{k+s})}{\Pi^{\beta}(\mathbf{X}^{k})}\frac{\mathcal{T}^{s}(\mathbf{X}^{k+s}\rightarrow\mathbf{X}^{k})}{\mathcal{T}^{s}(\mathbf{X}^{k}\rightarrow\mathbf{X}^{k+s})}\right\} ,\label{eq:asgen}\end{equation}
 $\mathcal{T}^{s}$ being the overall transition probability between
the two states. Let us first consider the case when $s=1$ and fix
the right direction $d=R$ (a similar derivation can be obtained for
$d=L$). In this case, the transition probability from the initial
state to the final state reads \begin{eqnarray}
\mathcal{T}^{1}(\mathbf{X}^{k}\rightarrow\mathbf{X}^{k+1}) & = & P(\mathbf{X}^{k},\rightarrow)\times R^{\tau}(x_{2M}^{k}\rightarrow x_{2M}^{k+1})\times\nonumber \\
 & \times & \left[1-K(\mathbf{X}^{k+1},\rightarrow)\right],\end{eqnarray}
 namely, it is the product of the probability of having chosen the
right direction, times the transition probability for the new tail
of the reptile, times the probability of stopping the updates after
one intermediate step. The inverse transition probability instead
reads \begin{eqnarray}
\mathcal{T}^{1}(\mathbf{X}^{k+1}\rightarrow\mathbf{X}^{k}) & = & P(\mathbf{X}^{k+1},\leftarrow)\times R^{\tau}(x_{0}^{k+1}\rightarrow x_{0}^{k})\times\nonumber \\
 & \times & \left[1-K(\mathbf{X}^{k},\leftarrow)\right],\end{eqnarray}
 which can be obtained reversing the time directions and considering
transitions from the head of the reptile instead that from the tail.
Therefore, the acceptance factor reads as \begin{eqnarray}
A^{1} & = & \min\left\{ 1,\frac{1-K(\mathbf{X}^{k},\leftarrow)}{P(\mathbf{X}^{k},\rightarrow)}\times\right.\nonumber \\
 &  & \left.\times\frac{w(x_{2M-1}^{k+1})}{w(x_{1}^{k})}\times\frac{P(\mathbf{X}^{k+1},\leftarrow)}{1-K(\mathbf{X}^{k+1},\rightarrow)}\right\} .\label{eq:a1}\end{eqnarray}
 For two intermediate transitions instead the transition probabilities
are \begin{eqnarray}
\mathcal{T}^{2}(\mathbf{X}^{k}\rightarrow\mathbf{X}^{k+2}) & = & P(\mathbf{X}^{k},\rightarrow)\times R^{\tau}(x_{2M}^{k}\rightarrow x_{2M}^{k+1})\times\nonumber \\
 & \times & K(\mathbf{X}^{k+1},\rightarrow)\times R^{\tau}(x_{2M}^{k+1}\rightarrow x_{2M}^{k+2})\times\nonumber \\
 & \times & \left[1-K(\mathbf{X}^{k+2},\rightarrow)\right],\end{eqnarray}
 and \begin{eqnarray}
\mathcal{T}^{2}(\mathbf{X}^{k+2}\rightarrow\mathbf{X}^{k}) & = & P(\mathbf{X}^{k+2},\leftarrow)\times R^{\tau}(x_{0}^{k+2}\rightarrow x_{0}^{k+1})\times\nonumber \\
 & \times & K(\mathbf{X}^{k+1},\leftarrow)\times R^{\tau}(x_{0}^{k+1}\rightarrow x_{0}^{k})\times\nonumber \\
 & \times & \left[1-K(\mathbf{X}^{k},\leftarrow)\right],\end{eqnarray}
 leading to the acceptance factor \begin{eqnarray}
A^{2}=\min\left\{ 1,\frac{1-K(\mathbf{X}^{k},\leftarrow)}{P(\mathbf{X}^{k},\rightarrow)}\times\frac{K(\mathbf{X}^{k+1},\leftarrow)}{K(\mathbf{X}^{k+1},\rightarrow)}\times\right.\nonumber \\
\left.\times\frac{w(x_{2M-1}^{k+1})}{w(x_{1}^{k+1})}\times\frac{P(\mathbf{X}^{k+2},\leftarrow)}{1-K(\mathbf{X}^{k+2},\rightarrow)}\times\frac{w(x_{2M-1}^{k+2})}{w(x_{1}^{k})}\right\}  & .\end{eqnarray}
 The generalization to generic $s$ intermediate steps is straightforward
and can be written as\begin{multline}
A^{s}=\min\left\{ 1,\frac{1-K(\mathbf{X}^{k},\leftarrow)}{P(\mathbf{X}^{k},\rightarrow)}\times\frac{P(\mathbf{X}^{k+s},\leftarrow)}{1-K(\mathbf{X}^{k+s},\rightarrow)}\times\right.\\
\left.\times\frac{w(x_{2M-1}^{k+s})}{w(x_{1}^{k})}\times\left[{\displaystyle \prod_{l=1}^{s-1}\frac{K(\mathbf{X}^{k+l},\leftarrow)}{K(\mathbf{X}^{k+l},\rightarrow)}\times\frac{w(x_{2M-1}^{k+l})}{w(x_{1}^{k+l})}}\right]\right\} .\label{eq:asexp}\end{multline}
 To find a simple solution for the unknown probabilities, we first
impose a cancellation for the intermediate acceptance factors, namely
\begin{equation}
\frac{K(\mathbf{X},\leftarrow)}{K(\mathbf{X},\rightarrow)}=\frac{w(x_{1})}{w(x_{2M-1})},\label{eq:Psatk0}\end{equation}
 this condition is satisfied by Eqs.~(\ref{eq:pkeeplefA}) and~(\ref{eq:pkeeprigA}).
Then, we notice that the acceptance factor can be written only in
terms of the final and the initial states as \begin{equation}
A^{s}=\min\left\{ 1,\frac{q(\mathbf{X}^{k+s},\leftarrow)}{q(\mathbf{X}^{k},\rightarrow)}\right\} .\label{eq:asfact}\end{equation}
 Further, we can impose the two factors $q$ to be independent on
the direction, i.e., the condition $q(\mathbf{X},\leftarrow)=q(\mathbf{X},\rightarrow)=q(\mathbf{X})$,
which is satisfied if \begin{eqnarray}
\frac{P(\mathbf{X},\leftarrow)}{1-K(\mathbf{X},\rightarrow)} & \times & w(x_{2M-1})=\nonumber \\
 & = & \frac{P(\mathbf{x},\rightarrow)}{1-K(\mathbf{x},\leftarrow)}\times w(x_{1}).\label{eq:Psatk}\end{eqnarray}
 Since the two time directions are mutually exclusive, it is also
true that $P(\mathbf{X},\leftarrow)+P(\mathbf{X},\rightarrow)=1$,
which allows us to solve Eq.~(\ref{eq:Psatk}) and obtain Eqs.~(\ref{eq:Pleft})
and~(\ref{eq:Prig}). The same reasoning can be repeated for the
left direction and, due to imposed homogeneity for the probabilities,
it can be checked that the detailed balance is satisfied for the left
direction too.

\section{Bounce algorithm, directed updates, and efficiency}

\label{sec:On-the-relationship}

In this Appendix we comment on the relationship between the directed-update
scheme and the bounce algorithm. If $\alpha=1$ is taken in Eqs.~(\ref{eq:pkeeplefA})
and~(\ref{eq:pkeeprigA}), then after $s$ updates along the direction
$d$, at the end of the Markov step $P(\mathbf{X}^{k+s},d)=0$, i.e.,
the next Markov step will be taken in the opposite direction, just
like the bounce algorithm. Although the two algorithms are similar
in this particular limit, there is an important difference which eventually
leads to a different computational efficiency. In order to elucidate
this point and to show the $\alpha$-dependence of the efficiency
of the directed updates, we have done a systematic comparison of the
two algorithms.

\begin{figure}
\includegraphics[width=1\columnwidth]{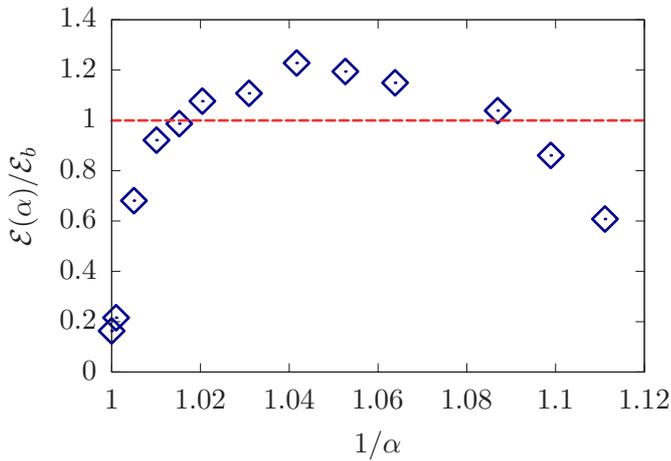} \caption{\label{fig:effic} Relative efficiency of the directed update scheme
and the bounce algorithm. The measured quantity is the ground-state
energy of the one-dimensional Heisenberg model on a chain of size
$L=80$ sites.}

\end{figure}

In particular, we have compared the efficiency of the directed updates
with the bounce algorithm for a one-dimensional Heisenberg model.
The computational efficiency is generally defined as \begin{equation}
\mathcal{E}=\frac{1}{\sigma_{O}^{2}T},\label{eq:eficc}\end{equation}
 where $\sigma_{O}^{2}$ is the square of the statistical error associated
to a given observable after a given computational time $T$. In Fig.~\ref{fig:effic},
we show the ratio between the directed-update scheme efficiency over
the bounce algorithm efficiency, for the measurement of the ground-state
energy of a one-dimensional chain.

%


We notice that the two sampling schemes have comparable performances,
being both based on a similar approach. As anticipated, it clearly
emerges from Fig.~\ref{fig:effic} that the two algorithms do not
have exactly the same behavior at $\alpha=1$, the maximum efficiency
of the directed updates being reached for lower values of $\alpha$.
This feature is due to the fact that when $\alpha$ is very close
to its saturation value, then a single Markov step can consist of
a conspicuous number of individual {}``sliding moves''. Even if
this situation leads to a fast decorrelation of configurations it
also leads to a rarefaction of the possibility to measure the desired
observables, which can eventually take place only at the end of the
Markov step and not during the individual moves. This leads to a worse
efficiency if compared to the bounce algorithm, where measurements
can be in principle done after every sliding move.

In conclusion, the performances of the two algorithms are very close,
although some advantages may arise from the use of the directed-updates.
We further notice that the purely Markovian approach introduced in
this paper could be slightly more efficient in cases where the number
of rejected configurations by the bounce algorithm is substantial
whereas {\em all} the generated configurations are accepted in
the directed update scheme.

\bibliographystyle{apsrev}

\end{document}